\documentclass[epj,twocolumn]{webofc}
\usepackage[varg]{txfonts}   
\usepackage{graphicx}
\usepackage{array}

\usepackage{soul}
\usepackage[usenames,dvipsnames]{xcolor}

\wocname{epj}
\woctitle{Seismology of the Sun and the Distant Stars 2016}
\begin{document}
\title{Deciphering the period spacing pattern in the oscillation spectrum of the SPB star KIC\,7760680}
%

\author{\firstname{Wojciech} \lastname{Szewczuk}\inst{1}\fnsep\thanks{\email{szewczuk@astro.uni.wroc.pl}} \and
        \firstname{Jadwiga} \lastname{Daszy\'nska-Daszkiewicz}\inst{1} \and
        \firstname{Przemys{\l}aw} \lastname{Walczak}\inst{1}
}

\institute{Astronomical Institute of the Wroc{\l}aw University, Kopernika 11, 51--622 Wroc{\l}aw, Poland
          }

\abstract{%
We present the analysis of KIC\,7760680, the rotating Slowly Pulsating B--type star
identified in the Kepler photometry. The oscillation spectrum of the star exhibits
a series of 36 frequencies which are quasi-equally spaced in period.
We confirm that
this series can be associated with prograde dipole modes of consecutive radial orders.
In our studies, the effects of rotation were included in the MESA equilibrium models as well as in the puslational calculations
in the framework of the traditional approximation.
We find that pulsational models computed with the OPLIB opacities best reproduce the observed frequency range.
The modified opacity data with an enhancement of the opacity at $\log T=5.3$, 5.46 and 5.06
were tested as well.
Increasing the OPLIB opacities by about 50\% at $\log T=5.3$ is sufficient to excite modes in the whole range of 36 frequency peaks of KIC\,7760680.
}
\maketitle
%
\section{Introduction}
\label{intro}
KIC\,7760680 was classified as the Slowly Pulsating B-type star by P{\'a}pics et al. \cite{Papics2015}
who found a series of 36 frequencies quasi-equally spaced in period from the Kepler photometry.
From spectroscopic observations the authors derived the effective temperature, $T_\mathrm{eff}=11650\pm 210$\,K,
surface gravity, $\log g=3.97\pm 0.08$, metallicity, $[M/H]=0.14\pm0.09$ (which correspond to about $Z=0.017$)
and the minimum value of the rotational velocity, $V_\mathrm{rot}\sin i=61.5\pm 5$ km\,s$^{-1}$.
Based on the zero-rotation asymptotic theory, P{\'a}pics et al. \cite{Papics2015} attributed
these frequencies to the prograde dipole gravity modes of high consecutive radial orders.
Recently, Moravveji et al. \citep{Moravveji2016} made a detailed seismic modelling of the star taking
into account effects of rotation in the framework of the traditional approximation
\cite[e.\,g.,][]{Lee_Saio1997,Townsend2003b,Townsend2003a,Townsend2005,2005MNRAS.364..573T, WD_JDD_AP2007,JDD_WD_AP2007}
applied to the non-rotating MESA \cite{Paxton2011,Paxton2013,Paxton2015}
evolutionary models. The authors found
that KIC\,7760680 is a moderately rotating star with the rotational frequency amounting to 26\% of its Roche break up frequency (0.4805 d$^{-1}$)
and with a mass $\sim 3.25 \mathrm M_\odot$. Furthermore, they constrained
the exponentially decaying convective core overshooting
parameter to $f_\mathrm{ov} \approx 0.024\pm 0.001$. Additionally, their best seismic
models employed extra diffusive mixing in the radiative envelope.
Here, we repeat the Moravveji et al. \citep{Moravveji2016} approach but our computations include the effects of rotation
also in equilibrium models. In Sect.\,2 we present results of our seismic modelling and conclusions ends the paper.

\section{Seismic modelling}

\begin{figure*}
\centering
\includegraphics[width=1.37\columnwidth,angle=270]{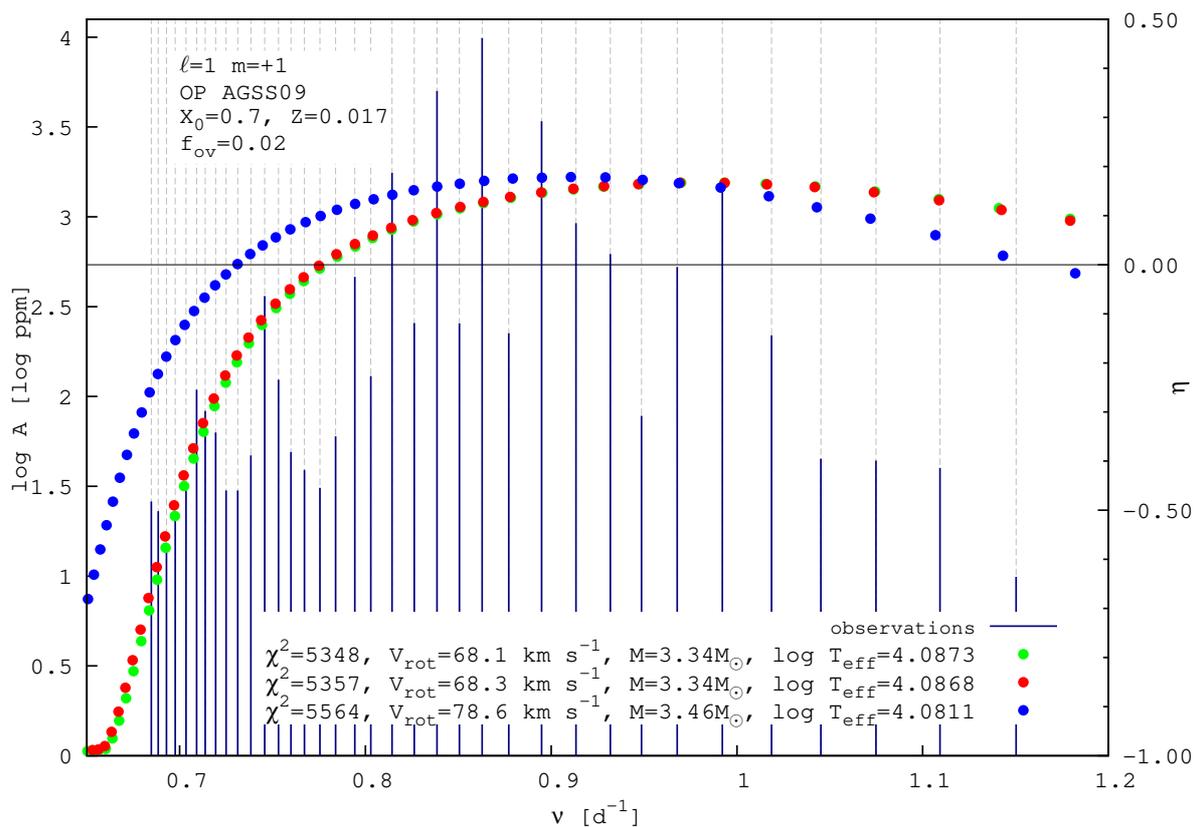}
\caption{The best seismic models calculated with OP opacities. The observed frequencies and their amplitudes
are plotted with vertical lines (left Y-axis).
Colour circles marked theoretical frequencies and their instability parameter, $\eta$, can bee reed
on the right Y-axis.}
\label{fig-1}       
\end{figure*}

\begin{figure*}
\centering
\includegraphics[width=1.37\columnwidth,angle=270]{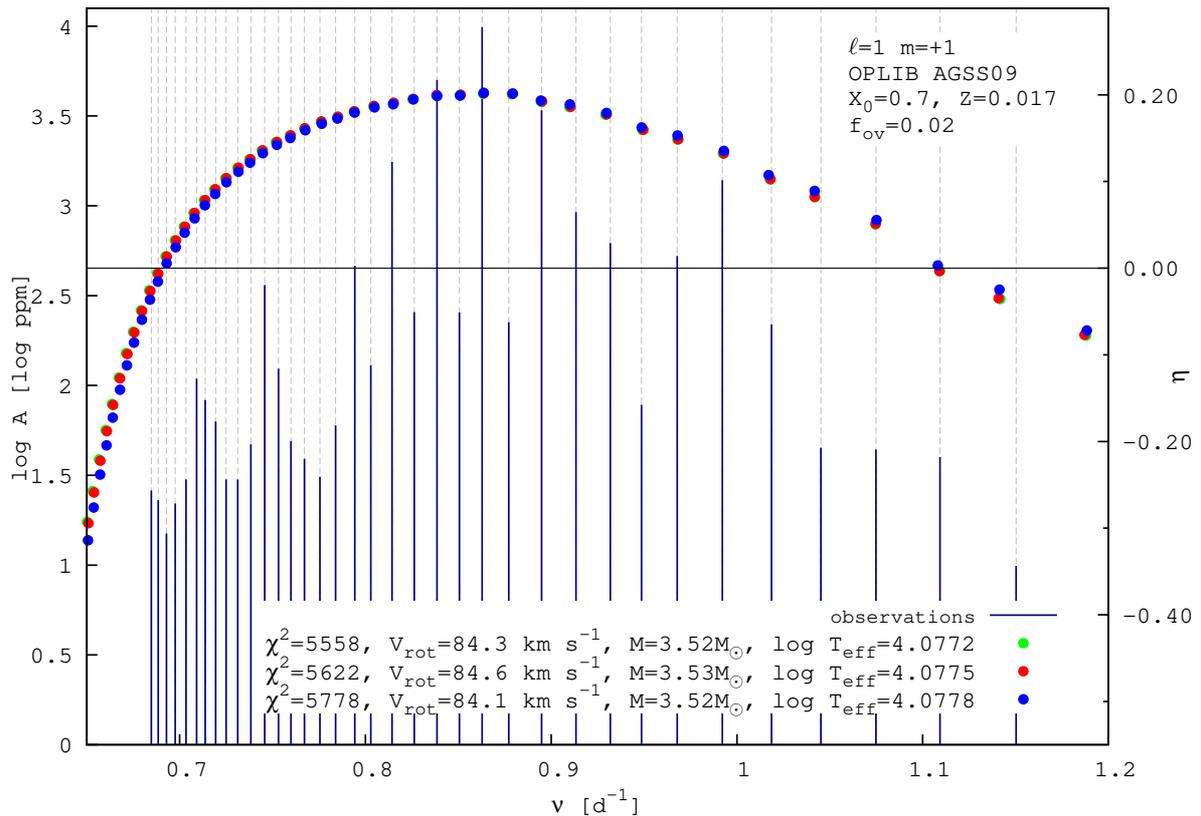}
\caption{The same as in Fig.\,\ref{fig-1} but OPLIB models are shown.}
\label{fig-2}       
\end{figure*}

\begin{figure*}
\centering
\includegraphics[width=1.37\columnwidth,angle=270]{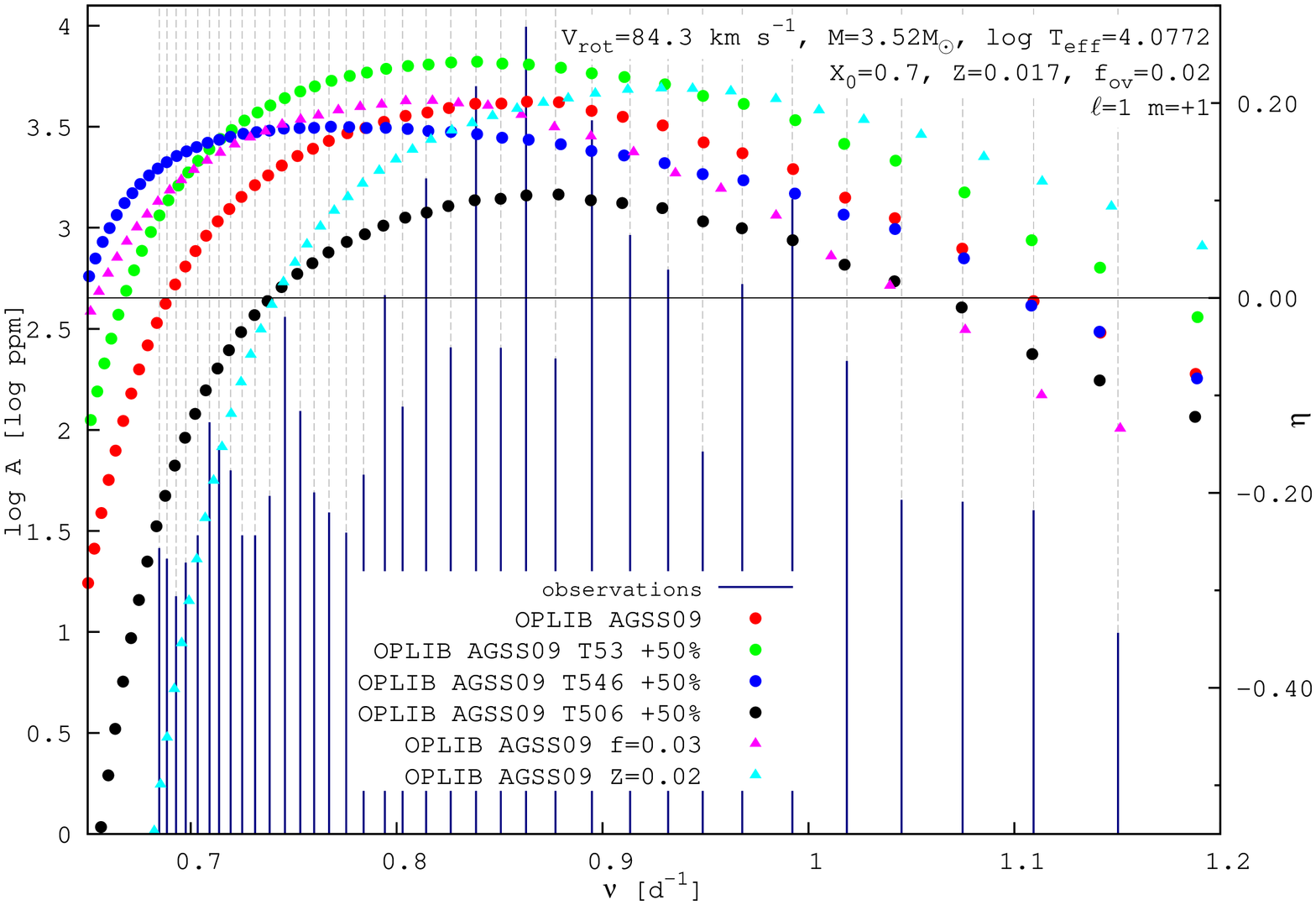}
\caption{Best OPLIB model recalculated with increased OPLIB opacities by 50\% at $\log T=5.30$
         (green circles), at $\log T=5.46$ (blue circles) and at $\log T=5.06$ (black circles)
         as well increased overshooting parameter (magenta triangles) and increased metallicity
         (cyan triangles). Original OPLIB model is marked by red circles.
         The coordinates meaning is the same as in Fig\,\ref{fig-1}.}
\label{fig-3}       
\end{figure*}

\label{sec-1}

Knowledge of the geometry of pulsational modes is essential in seismic modelling.
In the case of KIC\,7760680, the quasi-equally spaced in period frequency peaks suggest that these are
the modes with the same degree, $\ell$, azimuthal order, $m$, and consecutive radial orders, $n$.
According to P{\'a}pics et al \citep{Papics2015}, the frequencies are associated with
prograde dipole modes $\left(\ell=1,\,m=+1\right)$.

In order to check whether this statement is valid if the effects of rotation are included, both in the
equilibrium and pulsational calculations, we constructed a grid of rotating stellar models within
the $3\,\sigma$ error box in $T_\mathrm{eff}$ and $\log g$.
To this end we used  MESA code \cite{Paxton2011,Paxton2013,Paxton2015} considering a wide range
of the rotational velocity, from $V_\mathrm{rot}=62$ km\,s$^{-1}$ to the upper limit
of the validity of the traditional approximation.
As the upper limit we adopted 200 km\,s$^{-1}$ which is about a half of the  break-up velocity.
The step was $\Delta V_\mathrm{rot}=1$ km\,s$^{-1}$.
We included various rotational mixing mechanisms, i.e., dynamical shear instability, Solberg-H{\o}iland instability, secular shear instability,
Eddington-Sweet circulation and the Goldreich-Schubert-Fricke instability
\cite[see ][]{2000ApJ...528..368H,2005ApJ...626..350H,Paxton2013}.
We assumed the initial hydrogen abundance, $X_{0}=0.7$, metallicity $Z=0.017$, the chemical mixture
of Asplund et al. \cite{AGSS09}, and the overshooting parameter in exponential description, $f=0.02$.
Three commonly used opacity data were adopted: OP \cite{Seaton2005}, OPAL
\cite{OPAL1996} and OPLIB \cite{OPLIB2015,Colgan2016}.

Then, we calculated nonadiabatic pulsation models in the framework of traditional approximation
(here the step in $V_\mathrm{rot}$ was decreased to 0.1 km\,s$^{-1}$)
for modes which seemed to be good candidates to reproduce observed series of frequencies, i.e.,\,
(1,\,+0), (1,\,+1), (2,\,+0) and (2,\,+1).
Since for modes other than (1,\,+1) the best models give worse quality of the fit,
$\chi^{2}$ (defined as in Moravveji et al. \citep{Moravveji2016}),
by the order of magnitude, we confirm that the frequencies in the series
can be associated with the prograde dipole modes.

In Fig.\,1, the theoretical frequencies of (1,\,+1) modes for the best three OP models are
compared with the observed ones. The parameters of these models are summarised in Table\,\ref{tab-1}.
Without a doubt, the quality of fitting
needs  improvement. For the best models, we obtained the values of $\chi^2$ of the order of 5500 which is the value
almost three times higher than the one obtained by Moravveji et al. \citep{Moravveji2016}.
This is because in our calculations we fixed metallicity and overshooting parameter
compared to the cited authors. Moreover, Moravveji et al. fitted coefficient of the extra diffusive mixing in
the radiatively stable envelope whereas we employed rotational induced mixing with fixed coefficients.
In Fig.\,1 on the right Y-axis there is also shown the instability parameter $\eta$
($\eta\le0$ -- stable mode, $\eta>0$ -- unstable mode). As one can see modes with the lowest frequencies
in our best models are predicted to be stable. This is obviously contrary to the observations
and needs to be improved.

The quality of the fit is quantitatively the same for each considered opacity source
but for models with OPLIB opacities the theoretical instability domain is in an excellent agreement
with the observed frequencies. This is shown in Fig.\,\ref{fig-2}, in which three
best OPLIB models are plotted. As one can see only 3--4 modes are
marginally stable, $\eta\approx0$.
Parameters of these models as well as obtained with the OPAL
opacities are summarised in Table\,\ref{tab-1}.

Then, we recalculated the best OPLIB model with
different values of the metallicity and overshooting parameter
as well as with the modified OPLIB opacity profile.
The modifications of the OPLIB data consisted of an enhancement of the opacity at the depths corresponding to temperatures $\log T=5.3$, 5.46
and $\log T=5.06$. The increase of the opacity at $\log T=5.06$ mimics the new opacity bump identified in Kurucz models by Cugier \citep{Cugier2014}.
The modified models are shown in Fig.\,\ref{fig-3}.
The increase of the opacity at $\log T=5.3$ by about 50\% gives theoretical instability in the whole range of the observed frequencies.
Moreover, for each observed frequency we found a corresponding unstable mode. Increasing opacity at
$\log T=5.46$ helps in exciting low frequency modes whereas instability of high frequencies is unchanged. On the other
hand, adding the Kurucz bump results in reducing the pulsational instability, both, at low and high frequencies.
Larger value of the metallicity gives the instability domain shifted towards higher frequencies
whereas more effective overshooting
from the convective core acts in the opposite direction.

\begin{table}
\centering
\caption{The parameters of the best seismic models. Column 1 -- opacity data
        (\#1 -- OP, \#2 -- OPAL, \#3 -- OPLIB);
        Column 2 -- mass (in the solar mass units)        
        Column 3 -- logarithm of the effective temperature;
        Column 4 -- logarithm of the luminosity (in the solar luminosity units);
        Column 5 -- central hydrogen abundance; Column 6 -- rotational velocity (in km\,s$^{-1}$);
        Column 7 -- measure of the fitting quality.}
\label{tab-1}       
\begin{tabular}{ccccccc}
\hline
\vspace{-10pt}
   &     &                       &                              &                &                  &           \\
\vspace{1pt}
op & $M$ & $\log T_\mathrm{eff}$ & $\log \frac{L^{~}}{L_\odot}$ & $X_\mathrm{c}$ & $V_\mathrm{rot}$ & $\chi^2$  \\
\hline
\#1   &3.340& 4.0873                & 2.222           & 0.42          & 68.1             & 5348\\
\#1   &3.340& 4.0868                & 2.224           & 0.42          & 68.3             & 5357\\
\#1   &3.460& 4.0811                & 2.327           & 0.35          & 78.6             & 5564\\

\#2   &3.510& 4.0807                & 2.375           & 0.31          & 83.3             & 5786\\
\#2   &3.460& 4.0834                & 2.336           & 0.34          & 78.8             & 5851\\
\#2   &3.510& 4.0801                & 2.376           & 0.31          & 83.5             & 5998\\

\#3   &3.520& 4.0772                & 2.371           & 0.30          & 84.3             & 5558\\
\#3   &3.525& 4.0775                & 2.373           & 0.30          & 84.6             & 5622\\
\#3   &3.520& 4.0778                & 2.370           & 0.31          & 84.1             & 5778\\
\hline
\end{tabular}
\end{table}

\section{Conclusions}
\label{sec-con}
Our calculations confirmed P{\'a}pics et al. \citep{Papics2015} finding that 36
quasi-equidistant in period frequencies observed in KIC\,7760680 are most probably associated with the consecutive
prograde dipole modes. However, our best models have slightly higher masses
(from 3.340$\,\mathrm{M_\odot}$ to 3.525 $\,\mathrm{M_\odot}$ depending on the opacity data)
compared to 3.25$\,\mathrm{M_\odot}$ obtained by Moravveji et al. \citep{Moravveji2016}. The reason is that
we used rotating evolutionary models which give lower effective temperature for the same mass.
Rotational induced mixing can contribute to the differences as well. We got a little worse quality of the fit,
$\chi^{2} \sim 5500$, compared to the cited authors,
$\chi^{2} \sim 2000$ but in our calculations we have fixed the metallicity and the parameter of the core overshooting, which Moravveji et al. \citep{Moravveji2016} set as free.
It is prominent that instability domain of the (1,\,+1) modes obtained with the OPLIB data (see Fig.\,\ref{fig-2})
nearly cover the range of the observed 36 frequency peaks.
A slight increase of the opacity at $\log T=5.3$ gives an excellent agreement.
On the other hand, adding the Kurucz bump at $\log T=5.06$ spoils the agreement.
Finally, it should be added that  series of frequencies equidistant in period may be accidental as
it can be reproduced by modes of various angular numbers ($\ell,~m$; see e.g. \cite{Szewczuk2014}).
Therefore, it is desirable to confirm the results of mode identification
using another method, e.g., using multicolour time-series photometry.

\section*{\footnotesize Acknowledgements}


\footnotesize
This work was financially supported by the Polish National Science Centre grant 2015/17/B/ST9/02082.
Calculations have been carried out using
resources provided by Wroc{\l}aw Centre for Networking and Supercomputing (http://wcss.pl), grant no. 265.\\

%
%
%
%

\bibliography{Szewczuk}

\begin{thebibliography}{21}

\bibitem{Papics2015}
P.I. {P{\'a}pics}, A.~{Tkachenko}, C.~{Aerts}, T.~{Van Reeth}, K.~{De Smedt},
  M.~{Hillen}, R.~{{\O}stensen}, E.~{Moravveji}, Astrophys. J. \textbf{803},
  L25 (2015)

\bibitem{Moravveji2016}
E.~{Moravveji}, R.H.D. {Townsend}, C.~{Aerts}, S.~{Mathis}, Astrophys. J.
  \textbf{823}, 130 (2016)

\bibitem{Lee_Saio1997}
U.~{Lee}, H.~{Saio}, Astrophys. J. \textbf{491}, 839 (1997)

\bibitem{Townsend2003b}
R.H.D. {Townsend}, MNRAS \textbf{340}, 1020 (2003)

\bibitem{Townsend2003a}
R.H.D. {Townsend}, MNRAS \textbf{343}, 125 (2003)

\bibitem{Townsend2005}
R.H.D. {Townsend}, MNRAS \textbf{360}, 465 (2005)

\bibitem{2005MNRAS.364..573T}
R.H.D. {Townsend}, MNRAS \textbf{364}, 573 (2005)

\bibitem{WD_JDD_AP2007}
W.A. {Dziembowski}, J.~{Daszy{\'n}ska-Daszkiewicz}, A.A. {Pamyatnykh}, MNRAS
  \textbf{374}, 248 (2007)

\bibitem{JDD_WD_AP2007}
J.~{Daszynska-Daszkiewicz}, W.A. {Dziembowski}, A.A. {Pamyatnykh}, Acta Astron.
  \textbf{57}, 11 (2007)

\bibitem{Paxton2011}
B.~{Paxton}, L.~{Bildsten}, A.~{Dotter}, F.~{Herwig}, P.~{Lesaffre},
  F.~{Timmes}, Astrophys. J. Suppl. \textbf{192}, 3 (2011)

\bibitem{Paxton2013}
B.~{Paxton}, M.~{Cantiello}, P.~{Arras}, L.~{Bildsten}, E.F. {Brown},
  A.~{Dotter}, C.~{Mankovich}, M.H. {Montgomery}, D.~{Stello}, F.X. {Timmes}
  et~al., Astrophys. J. Suppl. \textbf{208}, 4 (2013)

\bibitem{Paxton2015}
B.~{Paxton}, P.~{Marchant}, J.~{Schwab}, E.B. {Bauer}, L.~{Bildsten},
  M.~{Cantiello}, L.~{Dessart}, R.~{Farmer}, H.~{Hu}, N.~{Langer} et~al.,
  Astrophys. J. Suppl. \textbf{220}, 15 (2015)

\bibitem{2000ApJ...528..368H}
A.~{Heger}, N.~{Langer}, S.E. {Woosley}, Astrophys. J. \textbf{528}, 368 (2000)

\bibitem{2005ApJ...626..350H}
A.~{Heger}, S.E. {Woosley}, H.C. {Spruit}, Astrophys. J. \textbf{626}, 350
  (2005)

\bibitem{AGSS09}
M.~{Asplund}, N.~{Grevesse}, A.J. {Sauval}, P.~{Scott}, Ann. Rev. Astron.
  Astrophys. \textbf{47}, 481 (2009)

\bibitem{Seaton2005}
M.J. {Seaton}, MNRAS \textbf{362}, L1 (2005)

\bibitem{OPAL1996}
C.A. {Iglesias}, F.J. {Rogers}, Astrophys. J. \textbf{464}, 943 (1996)

\bibitem{OPLIB2015}
J.~{Colgan}, D.P. {Kilcrease}, N.H. {Magee}, J.~{Abdallah}, M.E. {Sherrill},
  C.J. {Fontes}, P.~{Hakel}, H.L. {Zhang}, High Energy Density Physics
  \textbf{14}, 33 (2015)

\bibitem{Colgan2016}
J.~{Colgan}, D.P. {Kilcrease}, N.H. {Magee}, M.E. {Sherrill}, J.~{Abdallah},
  Jr., P.~{Hakel}, C.J. {Fontes}, J.A. {Guzik}, K.A. {Mussack}, Astrophys. J.
  \textbf{817}, 116 (2016)

\bibitem{Cugier2014}
H.~{Cugier}, Astron. \& Astrophys. \textbf{565}, A76 (2014)

\bibitem{Szewczuk2014}
W.~{Szewczuk}, J.~{Daszy{\'n}ska-Daszkiewicz}, W.~{Dziembowski},
  \emph{{Interpretation of the oscillation spectrum of HD 50230 - a failure of
  richness}}, in \emph{Precision Asteroseismology}, edited by J.A. {Guzik},
  W.J. {Chaplin}, G.~{Handler}, A.~{Pigulski} (2014), Vol. 301 of \emph{IAU
  Symposium}, pp. 109--112

\end{thebibliography}

\end{document}